\documentclass[11pt,letter]{IEEEtran}
\usepackage{amssymb}
\usepackage{amsmath}
\usepackage{graphicx}
\usepackage{cite}

\begin{document}

\title{On The Positive Definiteness of Polarity Coincidence
Correlation Coefficient Matrix}
\author{F.~Haddadi,~\IEEEmembership{Student~Member,~IEEE,}
        M.~M.~Nayebi,~\IEEEmembership{Senior~Member,~IEEE,}
        and~M.~R.~Aref
\thanks{This work was supported by Advanced Communication
Research Institute (ACRI), Sharif University of Technology,
Tehran, Iran. Authors are with the Department of Electrical
Engineering, Sharif University of Technology, Tehran, Iran
(e-mails: farzanhaddadi@yahoo.com, Nayebi@sharif.edu, and
Aref@sharif.edu).}}

\maketitle

\begin{abstract}
Polarity coincidence correlator (PCC), when used to estimate the
covariance matrix on an element-by-element basis, may not yield a
positive semi-definite (PSD) estimate. Devlin et al. \cite{Devlin},
claimed that element-wise PCC is not guaranteed to be PSD in
dimensions $p>3$ for real signals. However, no justification or
proof was available on this issue. In this letter, it is proved that
for real signals with $p\leq3$ and for complex signals with $p \leq
2$, a PSD estimate is guaranteed. Counterexamples are presented for
higher dimensions which yield invalid covariance estimates.
\end{abstract}



\IEEEpeerreviewmaketitle

\section{Introduction and Preliminaries}

\PARstart{P}{olarity} coincidence correlator (PCC) is a robust and
nonparametric estimator of bivariate correlation \cite{Devlin,
Wolff}. It is also a fast and low-cost estimator for applications
with extraordinary computational complexity. Radio astronomy is an
instance in which PCC is by far the most favorable correlator
\cite{Egau}.

Several researchers have investigated the statistical error of PCC
as an estimate of bivariate correlation \cite{Gabriel,Jacovitti92}.
In multivariate case, using PCC to estimate elements of the
covariance matrix does not guarantee a PSD matrix estimator
\cite{Devlin,Visuri}. Devlin et al. \cite[Sec. 4.4]{Devlin},
referring to a personal communication, claim that element-wise PCC
(or ''quadrant correlation''), may yield an invalid covariance
estimate for $p>3$ and real signals.

In this letter, we prove that for real signals with $p\leq3$, and
complex signals with $p\leq2$, PCC estimate is PSD. For higher
dimensions, counterexamples are presented which yield invalid
covariance estimates.

Let $x$ and $y$ be two zero-mean real random variables with
correlation coefficient $r$ distributed with elliptical symmetry. It
is well known that \cite{Visuri}:

\begin{equation}
\label{eq:def}
r=\sin\Big(\frac{\pi}{2}E\{\textrm{sgn}(x)\textrm{sgn}(y)\}\Big)
\end{equation}
where
\begin{equation}
\textrm{sgn}(x)= \left \{ \begin{array}{ll} +1 & \textrm{: $x\geq0$} \\
-1 & \textrm{: $x<0$}
\end{array} \right.
\end{equation}
Using (\ref{eq:def}), an estimate of $r$ from $N$ iid observations
$x_i , y_i , i=1,\dots,N$ is given by
\begin{equation}
\label{eq:est} \hat{r}=\sin\Big(\frac{\pi}{2}\frac{1}{N}\sum_{i=1}^N
s_{xi}s_{yi}\Big).
\end{equation}
where  $s_{xi}= \textrm{sgn}(x_i)$. In the complex case, we can
define the complex sign function as $\textrm{sgn}_c(x) \triangleq
\textrm{sgn}(\Re[x])+j\,\textrm{sgn} (\Im[x])$, where $\Re[x]$ and
$\Im[x]$ are real and imaginary parts of $x$, respectively. In the
Appendix, it is shown that
\begin{eqnarray}
\label{eq:cmp} \Re[r]=\sin\Big( \frac{\pi}{4}E\big\{ \Re \,
[\textrm{sgn}_c(x)\textrm{sgn}_c^*(y)]\big\}\Big)
\nonumber \\
\Im[r]=\sin\Big( \frac{\pi}{4}E\big\{ \Im \,
[\textrm{sgn}_c(x)\textrm{sgn}_c^*(y)]\big\}\Big)
\end{eqnarray}
where $(\cdot)^*$ denotes complex conjugate. Similar to
(\ref{eq:est}), an estimate for the complex case is obtained by
replacing expectation with the average as
\begin{eqnarray}
\label{eq:estcmp} \hat{r}_R=\sin\Big( \frac{\pi}{4} \frac{1}{N}
\sum_{i=1}^N  \, [s_{xiR}s_{yiR}+s_{xiI}s_{yiI}]  \Big)
\nonumber \\
\hat{r}_I=\sin\Big( \frac{\pi}{4} \frac{1}{N} \sum_{i=1}^N \,
[s_{xiI}s_{yiR}-s_{xiR}s_{yiI}]  \Big)
\end{eqnarray}
where $(\cdot)_R$ and $(\cdot)_I$ denote real and imaginary parts,
respectively.

\section{Main Result}

Let $\boldsymbol R_{p \times p}$ be the covariance matrix of $p$
random signals with unit diagonal elements and off-diagonal elements
$r_{ij} : i,j=1, \cdots ,p \,$. For $p=2$ case, a valid correlation
estimate should satisfy $|\hat{r}| \leq 1$. For the real case of
(\ref{eq:est}), $|\hat{r}| = |\sin (\cdot)| \leq 1$. For the complex
case, regarding (\ref{eq:estcmp}) define $\alpha$ and $\beta$ such
that $\hat{r}_R = \sin(\alpha)$ and $\hat{r}_I = \sin(\beta)$. Then
\begin{equation}
\label{eq:a+b} \alpha + \beta =  \frac{\pi}{4N} \sum_{i=1}^N \,
[s_{xiR}( s_{yiR} - s_{yiI}) + s_{xiI}( s_{yiR} + s_{yiI})]
\end{equation}
and it can be easily checked that the argument of summation in
(\ref{eq:a+b}) belongs to $\{\pm 2\}$. This yields $\alpha + \beta
\leq \frac{\pi}{2}$. In the same manner, we can show that $\pm \;
\alpha \, \pm \, \beta \leq \frac{\pi}{2}$ which gives $|\alpha| +
|\beta| \leq \frac{\pi}{2}$. Now it is straightforward to see that
$|\hat{r}|^2 = \sin^2(\alpha)+\sin^2(\beta) \leq
\sin^2(|\alpha|)+\sin^2(\frac{\pi}{2} - |\alpha|) = 1$.

For $p=3$ and real signals, we calculate the valid range of the
elements of a $3\times3$ covariance matrix. Then we show that PCC
estimate lies in this range.

\subsection{Valid Range of Covariance}

Let $\boldsymbol R \in \mathbb R^{3\times3}$ be a covariance matrix
with unit diagonal elements. Valid range of $r_{23}$ should be
calculated when $r_{12},r_{13} \in [-1,+1]$ are fixed. It can be
readily shown that $|\boldsymbol R|\geq 0$ implies that
\begin{eqnarray}
\label{eq:RngCnt} |\,r_{23} - r_{12}r_{13}| \leq \sqrt
{\big(1-r_{12}^2 \big) \big(1-r_{13}^2 \big)}\,.
\end{eqnarray}

\subsection{PCC Covariance Estimate}

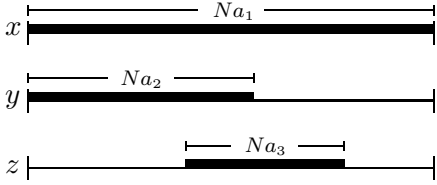
\begin{figure}[t]
\setlength{\unitlength}{3cm} \centering
    \begin{picture}(2,0.9)
        \multiput(0.1,0.1)(0,0.3){3}{\line(1,0){1.8}}
        \multiput(0.1,0.05)(0,0.3){3}{\line(0,1){0.1}}
        \multiput(1.9,0.05)(0,0.3){3}{\line(0,1){0.1}}
        \linethickness{1mm}
        \put(0.1,0.72){\line(1,0){1.8}}
        \put(0.1,0.42){\line(1,0){1.0}}
        \put(0.8,0.12){\line(1,0){0.7}}
        \linethickness{.1mm}
        \multiput(0.1,0.8)(1.05,0){2}{\line(1,0){0.75}}
        \multiput(0.1,0.78)(1.8,0){2}{\line(0,1){0.04}}
        \put(0.92,0.77){$\scriptstyle Na_1$}
        \multiput(0.1,0.5)(0.65,0){2}{\line(1,0){0.35}}
        \multiput(0.1,0.48)(1.0,0){2}{\line(0,1){0.04}}
        \put(0.51,0.47){$\scriptstyle Na_2$}
        \multiput(0.8,0.2)(0.5,0){2}{\line(1,0){0.2}}
        \multiput(0.8,0.18)(0.7,0){2}{\line(0,1){0.04}}
        \put(1.06,0.17){$\scriptstyle Na_3$}
        \put(0,0.69){$x$}
        \put(0,0.39){$y$}
        \put(0,0.08){$z$}
    \end{picture}
\caption{Polarity coincidence diagram of $\boldsymbol x,\boldsymbol
y,\boldsymbol z$. Black strips denote the packed positions of
polarity coincidences of each signal with signal $x$. The strips
lengths $Na_i$ are the number of polarity coincidences.}
\label{fig:model}
\end{figure}

Assume random sign sequences $s_x,s_y,s_z$ with length $N$. Consider
the positions of polarity coincidence with $s_x$ as black positions
or ''$+$'' and elsewhere as white or ''$-$''. Obviously all of the
positions in $s_x$ is ''$+$'' and $(s_{yi},s_{zi})$ have four states
of $\{ ++,+-,-+,-- \}$. Since the permutation of the samples does
not affect the estimate in (\ref{eq:est}), put the samples of
$s_x,s_y,s_z$ from left in the order of $\{ ++-,+++,+-+,+-- \}$ as
in Fig. \ref{fig:model}. Then any random sign sequences of $s_x$,
$s_y$ and $s_z$ can be replaced by the model in Fig. \ref{fig:model}
with appropriate strip lengths $Na_i$ (with $a_1=1$) and relative
positions of strips.

Let $\boldsymbol R_s$ be the covariance matrix of $\boldsymbol
s_x,\boldsymbol s_y,\boldsymbol s_z$ with elements $r_{sik} ,
i,k=1,2,3$. The maximum of $r_{s12}=+1$ occurs in $a_2=1$ and the
minimum of $r_{s12}=-1$ in $a_2=0$. In fact, $r_{s12}= \frac{1}{N}
\sum_{i=1}^{N}s_{xi}s_{yi}= \frac{1}{N}[Na_2 - (N-Na_2)] = 2a_2-1$,
in other words

\begin{equation}
a_i = \frac{1+r_{s1i}}{2}.
\end{equation}

$r_{s12}$ and $r_{s13}$ are determined by the values of $a_2$ and
$a_3$, $r_{sii}=1$, and the possible range of $r_{s23}$ should be
calculated. $r_{s23}$ depends on the number of polarity coincidences
of $\boldsymbol y$ and $\boldsymbol z$ which is maximum when the
strip of $\boldsymbol z$ is in the left corner, and minimum when it
is in the right corner. After some calculations, the range of
$r_{s23}$ is found as
\begin{equation}
\label{eq:RngSgn} |\,r_{s12}+r_{s13}|-1 \leq r_{s23} \leq
1-|\,r_{s12}-r_{s13}|.
\end{equation}

It should be noted that the effect of finite $N$ is the quantization
of the accessible values. Now, it can be readily verified that
\begin{eqnarray}
\sin \Big( \frac{\pi}{2} (1-|\,r_{s12}-r_{s13}|) \Big) =  \nonumber
\\ r_{12}r_{13} + \sqrt{\big(1-r_{12}^2 \big) \big(1-r_{13}^2
\big)}
\end{eqnarray}
and
\begin{eqnarray}
\sin \Big( \frac{\pi}{2} (\,|\,r_{s12}+r_{s13}|-1) \Big) = \nonumber
\\ r_{12}r_{13} - \sqrt{\big(1-r_{12}^2 \big) \big(1-r_{13}^2
\big)}.
\end{eqnarray}
Therefore, $\hat{r}_{23}=\sin \big( \frac{\pi}{2} \, r_{s23} \big)$
satisfies (\ref{eq:RngCnt}). This, besides $|\hat{r}_{12}| < 1 $ and
$|\hat{r}_{13}| < 1 $ can be used to show that $|\hat{\boldsymbol
R}| \geq 0$ (as in (\ref{eq:RngCnt})) and the assertion is proved
that for $p=3$ and real data, PCC estimate is a valid covariance
matrix.

\section{Counterexamples}

In this section, some counterexamples are presented to show that PCC
covariance estimate is not guaranteed to be PSD in dimensions $p>3$
for real signals and $p>2$ for complex signals. In real data case
with $p=4$ and number of observations $N=4$, the real sign sequences
in Table \ref{Table} results in an invalid covariance estimate.
After simple computations, we will have $r_{s12} = r_{s34}=0$ and
$r_{s13} = r_{s14}= r_{s23} = r_{s24}= 0.5$. The covariance estimate
will be
\begin{table}
    \renewcommand{\arraystretch}{1.5}
    \caption{Counterexamples for real and complex data} \label{Table}
        \centering
        \begin{tabular}{c||c}
            \hline
            \bfseries Real Case & \bfseries Complex Case \\
            \hline
            \hline
            \renewcommand{\arraystretch}{1.1}
            \begin{tabular}{c|c}
                $s_x$ & $+$ $+$ $+$ $+$ \\
                \hline
                $s_y$ & $+$ $+$ $-$ $-$ \\
                \hline
                $s_z$ & $+$ $+$ $+$ $-$ \\
                \hline
                $s_w$ & $+$ $+$ $-$ $+$ \\
            \end{tabular}
            &
            \renewcommand{\arraystretch}{1.48}
            \begin{tabular}{c|c}
                $s_x$ & $++$ \quad $++$ \\
                \hline
                $s_y$ & $++$ \quad $-+$ \\
                \hline
                $s_z$ & $++$ \quad $--$ \\
           \end{tabular} \\
           \hline
           \hline
        \end{tabular}
\end{table}
\begin{displaymath}
\hat{\boldsymbol R}_1 = \left[ \begin{array}{cccc} 1 & 0 & 0.7 & 0.7 \\ 0 & 1 & 0.7 & 0.7 \\
0.7 & 0.7 & 1 & 0 \\ 0.7 & 0.7 & 0 & 1 \\
\end{array} \right]
\end{displaymath}
with eigenvalues $[-0.4,1,1,2.4\,]$. Then $\hat{\boldsymbol R}_1$,
with a negative eigenvalue, is not a valid covariance matrix.

We can augment this example to give a counterexample for dimension
$p=5$. Repeat each sign twice to have four signals with number of
observations $2N$. Note that the covariance matrix does not
change. Now, add a new signal with alternating sign in each
sample. The covariance estimate will be

\begin{displaymath}
\hat{\boldsymbol R}_{aug} = \left[ \begin{array}{cc}
\hat{\boldsymbol R}_1 & \boldsymbol 0 \\  \boldsymbol 0^T & 1 \\
\end{array} \right].
\end{displaymath}
where $\boldsymbol 0$ is the $4 \times 1$ vector of zeros. As a
consequence of the structure of $\hat{\boldsymbol R}_{aug}$,
eigenvalues of $\hat{\boldsymbol R}_1$ are also eigenvalues of
$\hat{\boldsymbol R}_{aug}$. Therefore, $\hat{\boldsymbol R}_{aug}$
is an invalid covariance matrix. This procedure can continue to
produce counterexamples for higher dimensions in real data case.

In case of complex signals, $p=3$ and $N=2$, a counterexample is
given in Table \ref{Table}, where ''$-+$'' denotes $-1+j$. The
resulting estimate is
\begin{displaymath}
\hat{\boldsymbol R} = \left[ \begin{array}{ccc} 1 & 0.7-j\,0.7 & 0 \\
0.7+j\,0.7 & 1 & 0.7-j\,0.7 \\ 0 & 0.7+j\,0.7 & 1 \\
\end{array} \right]
\end{displaymath}
with eigenvalues $[-0.4,1,2.4\,]$ which make $\hat{R}$ an invalid
covariance matrix. Augmentation of the complex signal set for
higher dimensions is similar to the real case, except that the new
added signal alternates between ''$++$'' and ''$--$''.

\section{Applications of the Results}

In this section, we discuss the practical usefulness of the main
results of this letter which focuses on low number of sensors. In
the signal processing context, covariance estimation often arises
in the multi-sensor applications where parameters of interest are
functions of the true data covariance matrix. Although PCC
estimate of the covariance matrix exhibits attractive features
such as robustness and extremely low complexity, it cannot be
guaranteed to be PSD in the applications with large number of
sensors.

Selection of the number of sensors in an application depends on both
nature of the problem and practical limitations. In theory, more
sensors always results in a better estimate, as proved in many cases
such as direction finding through examination of the Cramer-Rao
bounds \cite{stoica}. In practice, complexity issues usually limit
the number of sensors. Large arrays are used whenever performance be
of the main importance regardless of the cost. In such cases as DOA
estimation in military environments (radar and sonar), thousands of
sensors are not uncommon. Nevertheless, most low-cost civil
applications use very few sensors. In the following, we consider
some of these applications.

\subsection{MIMO Communication Systems}

Multiple antenna systems are an integral part of the most new
wireless communication systems increasing user and data capacity
(e.g. UMTS/W-CDMA, 802.11n WLAN, 60 GHz WPAN). Multiple antennas
can provide diversity gain and/or better antenna gain through
beamforming in base station and/or handset. Beamformers (e.g.
conventional or Capon) usually utilize an estimate of the array
covariance matrix \cite{Vanveen}, that may be obtained using PCC
as a power-saving estimator. It is well known that performance
improvement due to diversity gain reduces as the number of
antennas increases. This, besides space limit on the handset and
coupling phenomena have resulted in the prevalence of MIMO systems
with very few (usually 2 to 4) antennas \cite{Browne, Hui}.

\subsection{Blind Source Separation (BSS)}

BSS has found numerous potential applications in the field of
audio signal processing \cite{Torkkola}. An array of microphones
is used to gather multiple signal mixtures and diverse methods are
used to extract signals from these observations. A large class of
BSS methods use real-valued inter-sensor covariances with
different time lags to estimate the mixing matrix and desired
signals (e.g. SOBI \cite{Belouchrani}, JADE \cite{Cardoso}). This
also includes input signals whitening as a preprocessing that
converts the convolutive source separation problem to a simpler
independent component analysis (ICA) problem. This family of
two-step algorithms is known as AMUSE (Algorithm for MUltiple
Source Extraction). PCC, as a fast correlator, can make real-time
operation more feasible in these methods. For realistic situations
where we have fewer sensors than sources, underdetermined methods
are proposed  \cite{Winter}. Many methods are presented for the
special case of 2 sensors and multiple sources (e.g. DUET
\cite{Yilmaz}, and \cite{Comon}), and also quite few sensors are
common to many realizations of the methods \cite{Belouchrani,
Winter}.

\appendix[Complex PCC]
Let $x,y$ be two zero-mean, unit-variance, and circularly symmetric
complex random variables with independent real and imaginary parts.
To prove (\ref{eq:cmp}), we expand the expectation as
\begin{eqnarray}
\label{eq:appexp}
E \{ \textrm{sgn}_c(x) \textrm{sgn}_c^*(y) \} =
E\big\{ [ \textrm{sgn}( x_Ry_R) + \textrm{sgn}( x_Iy_I) ] \nonumber
\\ + j\,[\,\textrm{sgn}( x_Iy_R) - \textrm{sgn}( x_Ry_I)] \big\}.
\end{eqnarray}
Furthermore, $E\{ xy^* \} = r$ implies that
\begin{eqnarray}
E\{ x_R y_R + x_I y_I \} = r_R \nonumber \\
E\{ x_Iy_R - x_R y_I \} = r_I.
\end{eqnarray}
Circular symmetry of $x$ and $y$ yields
\begin{eqnarray}
E\{ x_R y_R \} = E\{ x_I y_I \} = r_R/\,2 \nonumber \\
E\{ x_Iy_R\} = -E\{ x_R y_I \} = r_I/\,2
\end{eqnarray}
and $E\{ x_R ^2 \} = E\{ x_I^2 \} = E\{ y_R ^2 \} = E\{ y_I^2 \} =
\frac{1}{2}\,$. Then the correlation coefficients will be
\begin{eqnarray}
\label{eq:appcov}
\textrm{Cor}\,( x_R,y_R ) = \textrm{Cor}\,( x_I,y_I ) = r_R \nonumber \\
\textrm{Cor}\,( x_I,y_R ) =-\,\textrm{Cor}\,( x_R,y_I ) = r_I.
\end{eqnarray}
Substituting (\ref{eq:appcov}) and (\ref{eq:def}) into
(\ref{eq:appexp}) gives
\begin{eqnarray}
E \{ \textrm{sgn}_c(x) \textrm{sgn}_c^*(y) \} = 4/\pi\, [ \,
\sin^{-1} (r_R) + j \sin^{-1} (r_I)\, ]
\end{eqnarray}
which implies (\ref{eq:cmp}).


\end{document}